\documentclass[prl,twocolumn,showpacs,preprintnumbers,amsmath,amssymb]{revtex4}

\usepackage{graphicx}
\usepackage{dcolumn}
\usepackage{bm}

\begin{document}


\title{Gaussian approximation potentials: The accuracy of quantum mechanics, without the electrons}

\author{Albert P. Bart\'ok}
\affiliation{%
Cavendish Laboratory, University of Cambridge, J J Thomson Avenue, Cambridge, CB3 0HE, UK}
\author{Mike C. Payne}%
\affiliation{%
Cavendish Laboratory, University of Cambridge, J J Thomson Avenue, Cambridge, CB3 0HE, UK}
\author{Risi Kondor}
\affiliation{Center for the Mathematics of Information, California Institute of Technology, MC 305-16, Pasadena, CA 91125, USA}
\author{G\'abor Cs\'anyi}
\affiliation{
Engineering Laboratory, University of Cambridge, Trumpington Street, Cambridge, CB2 1PZ, UK}%

\date{\today}

\begin{abstract}
We introduce a class of interatomic potential models that can be automatically generated from data consisting of the energies and forces experienced by atoms, as derived from quantum mechanical calculations. The models do not have a fixed functional form and hence are capable of modeling complex potential energy landscapes. They are systematically improvable with more data. We apply the method to bulk crystals, and test it by calculating properties at high temperatures. Using the interatomic potential to generate the long molecular dynamics trajectories required for such calculations saves orders of magnitude in computational cost.
\end{abstract}

\maketitle

\begin{figure}
\includegraphics[width=8cm]{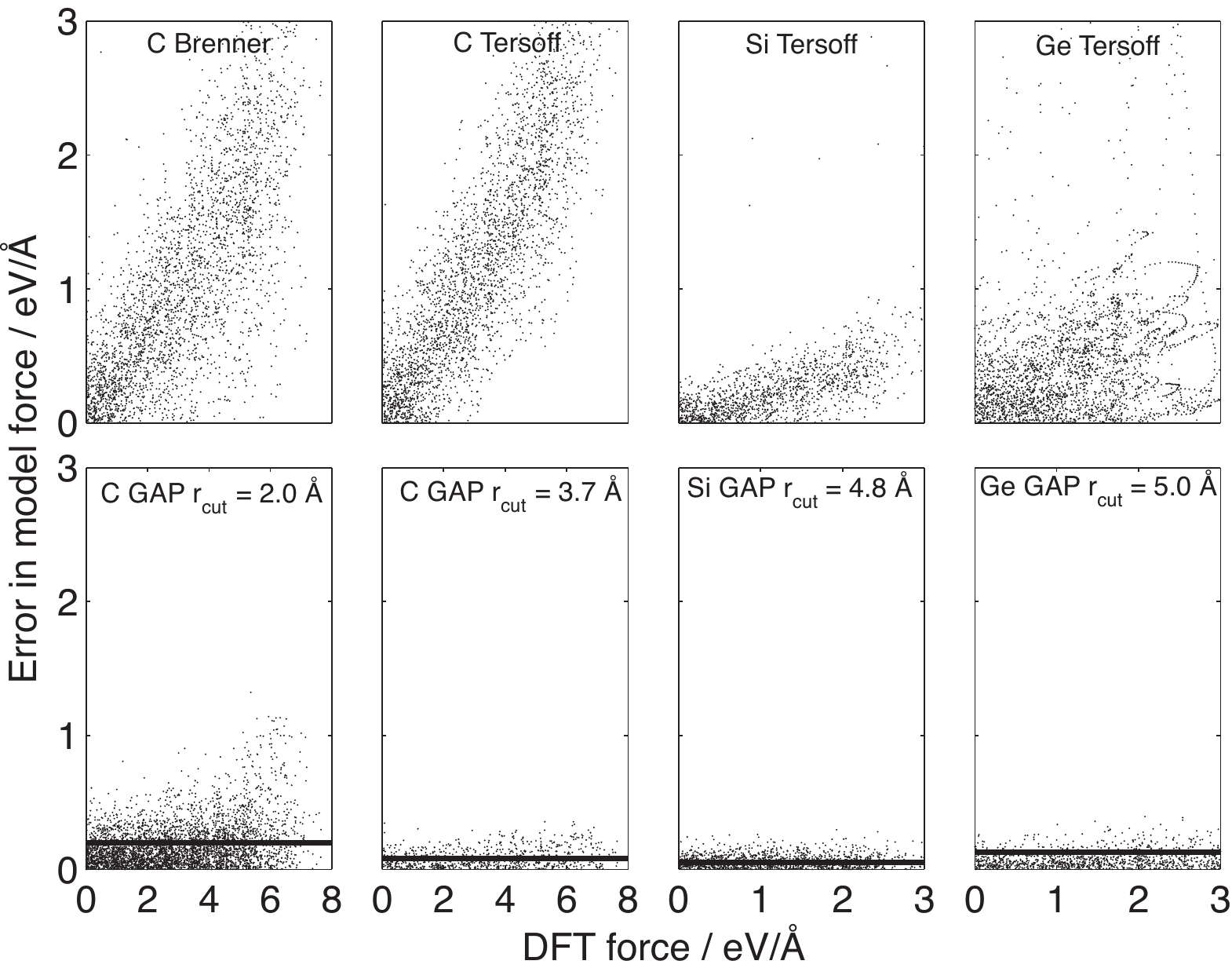}
\caption{\label{fig:forcecorr} Deviation of atomic forces between DFT and various models:  the Brenner \cite{Brenner03} and Tersoff\cite{TER02} potentials, and different GAP models for different semiconductors. In the bottom row the horizontal lines corresponds to the smallest standard deviation of the error theoretically attainable given the range of  the potential (see text). %
The configurations are taken from molecular dynamics  runs at 1000~K.}
\end{figure}

Atomic scale modeling of materials is now routinely and widely applied, and encompasses a range of techniques from exact quantum chemical methods \cite{szabobook} through density functional theory (DFT) \cite{dftoverview01} and semi-empirical quantum mechanics \cite{finnisbook} to analytic interatomic potentials \cite{brenner02}. The associated trade-offs in accuracy and
computational cost are well known. Arguably, there is a gap between models that treat electrons
explicitly, and those that do not. Models in the former class are in practice limited to handling a few thousand atoms, while the simple analytic interatomic potentials are limited in accuracy, regardless of how they are parametrized. The panels in the top row of Fig.~\ref{fig:forcecorr} illustrates the typical performance of analytic potentials in bulk semiconductors. Perhaps surprisingly, potentials that are generally regarded as adequate for describing these bulk phases show significant deviation from the quantum mechanical potential energy surface. This in turn gives rise to significant errors in predicting properties such as elastic constants and phonon spectra.

In this letter we are concerned with the problem of modeling the Born-Oppenheimer potential energy surface (PES) of a set of atoms, but without recourse to simulating the electrons explicitly. We mostly restrict our attention to modeling the bulk phases of carbon, silicon, germanium, iron and gallium nitride, using a unified framework. Even such single-phase potentials could be useful for calculating physical properties, e.g. the thermal expansion coefficient, the phonon contribution to the thermal conductivity, the temperature dependence of the phonon modes, or as part of QM/MM hybrid schemes \cite{ROP}. 

The first key insight is that this is actually practicable: the reason that
interatomic potentials are at all useful is that the PES is a relatively smooth function of the
nuclear coordinates.  Improving potential modeling is difficult {\em not} because the PES is rough,
but because it does not easily decompose into simple closed functional
forms. Secondly, away from isolated quantum critical points, the
behavior of atoms is localized in the sense that if the total energy of a system is written as a
sum of atomic energies,
\begin{equation}
E = \sum_i^\textrm{Atoms} \varepsilon(\{\mathbf{r}_{ij}\}),
\label{ip}
\end{equation}
where $\mathbf{r}_{ij} = \mathbf{r}_j-\mathbf{r}_i$ is the
relative position between atoms $i$ and $j$, then good approximations of $E$ can be
obtained by restricting the set of
atoms over which the index $j$ runs to some fixed neighborhood of atom $i$, i.e.,  $r_{ij} <
r_\textrm{cut}$. In fact, we take Eq.~(\ref{ip}) with this restriction as the defining feature of an
interatomic potential. Note that in general it is desirable to separate out Coulomb and dispersion
terms from the {\em atomic energy function}, $\varepsilon(\{\mathbf{r}_{ij}\})$, because the covalent part that remains can then be
localized much better for the same overall accuracy.  The strict localization of $\varepsilon$
enables the independent computation of atomic energies. However, it also puts a limit on the accuracy with which the PES
can be approximated.  Consider an atom whose environment {\em inside} $r_\textrm{cut}$ is fixed. The
true quantum mechanical force on this atom will show a variation, depending on its environment {\em outside} the
cutoff. An estimate of this variance is shown  on Fig.~\ref{fig:forcecorr} by the horizontal lines:
no interatomic potential with the given cutoff can have a lower typical force error.

To date, two works have attempted to model the PES in its full generality. In the first
\cite{Wang09}, small molecules were modeled by expanding the total energy in polynomials of all the
atomic coordinates, without restricting the range of the atomic energy function. While this gave
extremely accurate results, it cannot scale to more than a few atoms. More recently, a neural
network was used to model the atomic energy \cite{neur01}. Our philosophy and aims are similar to
the latter work: we compute $\varepsilon(\{\mathbf{r}_{ij}\})$ by interpolating a set of stored {\em
reference} quantum mechanical results using a carefully constructed measure of similarity between
{\em atomic neighborhoods}. We strive for computational efficiency in our use of expensive {\em ab
initio} data by using both the total energy and the atomic forces to obtain the best possible
estimate for $\varepsilon$ given our assumptions about its smoothness. Furthermore, our scheme makes
the generation of potential models automatic, with almost no need for human intervention in going
from quantum mechanical data to the final interatomic potential model.  In
the following, we present an overview of our formalism. Detailed derivations are given in the
Supplementary Information (SI).

The atomic energy function is invariant under translation, rotation and the permutation of atoms.
One of the key ideas in the present work is to represent atomic neighborhoods in a transformed
system of coordinates that accounts for these symmetries.  Ideally, this mapping should be
one-to-one: mapping different neighborhood configurations to the same coordinates would 
introduce systematic errors into the model that cannot be improved by adding more quantum mechanical
data.  We begin by forming a local atomic density from the neighbors of atom $i$, as
\begin{equation}
\rho_i (\mathbf{r})= \delta(\mathbf{r}) + \sum_j \delta(\mathbf{r}-\mathbf{r}_{ij}) f_\textrm{cut}(|\mathbf{r}_{ij}|),
\label{eq:rho}
\end{equation}
where $f_\textrm{cut}(r) = 1/2 + \cos(\pi r / r_\textrm{cut})/2$ is a cutoff function, in which the
cutoff radius $r_\textrm{cut}$ reflects the spatial scale of the interactions.
The choice of cutoff function is somewhat ad-hoc: any smooth function with compact support could be used.

The local atomic density is invariant to permuting the atoms in the neighborhood.  One way
to achieve rotational invariance as well would be to expand it in spherical harmonics and a set of radial basis functions and appropriately combine the resulting coefficients,  similarly to how the structure factor is computed
from Fourier components. However, just as the structure factor (a two-point correlation) is missing
all ``phase'' information (the relative phases of the different plane waves), such a set of
spherical invariants would lose a lot of information about the configuration of the neighborhood. 
In contrast, the {\em bispectrum} \cite{bis01}, which is a three-point correlation function, is a
much richer system of invariants, and can provide an almost one-to-one representation of the atomic
neighborhood.

In our method we first project the atomic density onto the surface of the four-dimensional unit
sphere, similarly to how the Riemann sphere is constructed, with the transformation
\begin{equation}
\mathbf{r}\equiv\left(\begin{matrix}x\\y\\z\end{matrix}\right) \rightarrow
\begin{array}{lcl}\phi &=& \arctan(y/x) \\ \theta &=& \arccos(z/|\mathbf{r}|) \\ \theta_0 &=& |\mathbf{r}|/r_0 \end{array}
\end{equation}
where $r_0 > r_\textrm{cut}/\pi$. The advantage of this is that the 4D surface contains all the
information from the 3D spherical region inside the cutoff,  including the radial dimension, and
thus 4D spherical harmonics (also called Wigner matrices, $U^j_{m'm}$\cite{sphericalH_4D_book})
constitute a natural complete basis for the interior of the 3D sphere, without the need for radial basis functions. The projection of the atomic density
on the surface of the 4D sphere can therefore be expanded in 4D spherical harmonics using
coefficients (dropping the atomic index $i$ for clarity)
\begin{equation}
c^j_{m'm} = \langle U^j_{m'm} | \rho \rangle
\end{equation}
The bispectrum built from these coefficients is given by 
\begin{equation}
\begin{split}
B_{j_1,j_2,j} = \sum_{m'_1, m_1 = -j_1}^{j_1} \sum_{m'_2, m_2 = -j_2}^{j_2}
\sum_{m', m = -j}^{j}
\bigl( c^j_{m'm} \bigr)^* \\C^{jm}_{j_1 m_1 j_2 m_2} C^{jm'}_{j_1 m'_1 j_2 m'_2}
c^{j_1}_{m'_1 m_1} c^{j_2}_{m'_2 m_2}
\end{split}
\end{equation}
where $C^{jm}_{j_1 m_1 j_2 m_2}$ are the ordinary Clebsch-Gordan coefficients. The elements of this three-index array, which we will denote by
$\mathbf{b}_i$ for atom $i$, are invariant with respect to permutation of atoms and rotations of 4D
space, and hence also 3D space. In practice, we use only a truncated version, with $j,j_1,j_2 \le
J_\textrm{max}$, corresponding to a limit in the spatial
resolution with which we describe the atomic neighborhood.


Determining the PES is now reduced to interpolating the atomic energy in the
truncated bispectrum-space, and for this we use a non-parametric method called Gaussian Process (GP)
regression \cite{RW0,info01}. 
In the GP framework, assuming Gaussian basis functions, 
the best estimate for the atomic energy function is given by 
\begin{equation}
\varepsilon(\mathbf{b}) = \sum_n \alpha_n e^{-\frac12 \sum_l \left[(b_{l}-b_{n,l})/\theta_l\right]^2} \equiv \sum_n \alpha_n G(\mathbf{b}, \mathbf{b}_n),
\label{eq:GP}
\end{equation}
where $n$ and $l$ range over the reference configurations and bispectrum components, respectively and  $\{\theta_l\}$ are (hyper)parameters. The GP is called a non-parametric method because the
kernels $G$ are not fixed but centered on the data, and hence, loosely, any continuous function can
in principle be obtained from Eq.~(\ref{eq:GP}) \cite{steinwart_machine_learning01}. 

The GP differs from a simple radial basis function least-squares fit in the way the coefficients
$\alpha_n$ are computed. The {\em covariance}, i.e. the measure of similarity, of the reference configurations is
defined as 
\begin{equation}
C_{n n^\prime} = \delta^2 G(\mathbf{b}, \mathbf{b}^{\prime})+ \sigma^2 \mathbf{I}
\label{eq:C}
\end{equation} 
where $\delta$ and $\sigma$ are two further hyperparameters and $\mathbf{I}$ is the identity matrix. The interpolation coefficients are then given by
\begin{equation}
\{\alpha_n\} \equiv \bm{\alpha} = \mathbf{C}^{-1} \mathbf{y},
\label{eq:alpha}
\end{equation}
where $\mathbf{y} = \{y_n\}$ is the set of reference values (quantum mechanical energies). This simple expression for 
the coefficients is derived in detail in \cite{info01}.  Thus Eq.~(\ref{eq:GP}) gives the atomic energy function in closed form as a function of the quantum mechanical data. 

In addition to preserving exact symmetries, another hurdle is that although we wish to infer the
atomic energy function, the data we can collect directly are not values of atomic energies, but
total energies of sets of atoms, and forces on atoms, the latter being sums of partial derivatives
of neighboring atomic energies\cite{sebastian_thesis}. Furthermore, our data will be heavily correlated: e.g. the neighborhoods
of atoms in a slightly perturbed ideal crystal are very similar to each other. Both of these
problems are solved by applying a sparsification procedure\cite{spgp01}, in which a predetermined
number (much smaller than the total data size) of ``sparse'' configurations are chosen
randomly from the set of all configurations and the data values $\mathbf{y}$ in Eq.~(\ref{eq:alpha})
are replaced by linear combinations of all data values. The models in this work used 300 such sparse configurations. The final expression for the model, which we call GAP, is derived in the SI.  

All the DFT data in this work was generated with the Castep package\cite{castep01}.
The reference
configurations were obtained by randomly displacing the atoms and the lattice vectors from their
equilibrium values in 2, 8, 16 and 64-atom cubic unit cells by up to $0.2~\textrm{\AA}$.  


\begin{table}
\begin{tabular}{rrrcc}
\vrule&DFT&GAP&Brenner&Tersoff (T)\\
\hline
1x1 unreconstructed \vrule&6.41&6.36&4.46&2.85\\
2x1 Pandey \vrule&4.23&4.40&3.42&4.77\\
\end{tabular}
\vskip\baselineskip
\begin{tabular}{rrrrcrrrcrrr}
\vrule&\multispan{3}\hfill C\hfill&\vrule&\multispan{3}\hfill Si\hfill&\vrule&\multispan{3}\hfill Ge\hfill\\
\vrule&DFT&GAP&T&\vrule&DFT&GAP&T&\vrule&DFT&GAP&T\\
\hline
$C_{11}$ \vrule& 1118 & 1081  &1072&\vrule&154&152&143&\vrule&108&114&138\\
$C_{12}$ \vrule& 151 & 157   &108&\vrule&56&59&75&\vrule&38&35&44\\
$C_{44}^0$ \vrule& 610 & 608   &673&\vrule&100&101&119&\vrule&75&75&93\\
$C_{44}$ \vrule& 603 & 601  &641&\vrule&75&69&69&\vrule&58&54&66\\
\end{tabular}
\caption{\label{tab:elastic} Table of relaxed diamond surface energies in J/m$^2$(top) and elastic constants, in units of GPa (bottom).}

\end{table}

\begin{figure}
\includegraphics[width=8.5 cm]{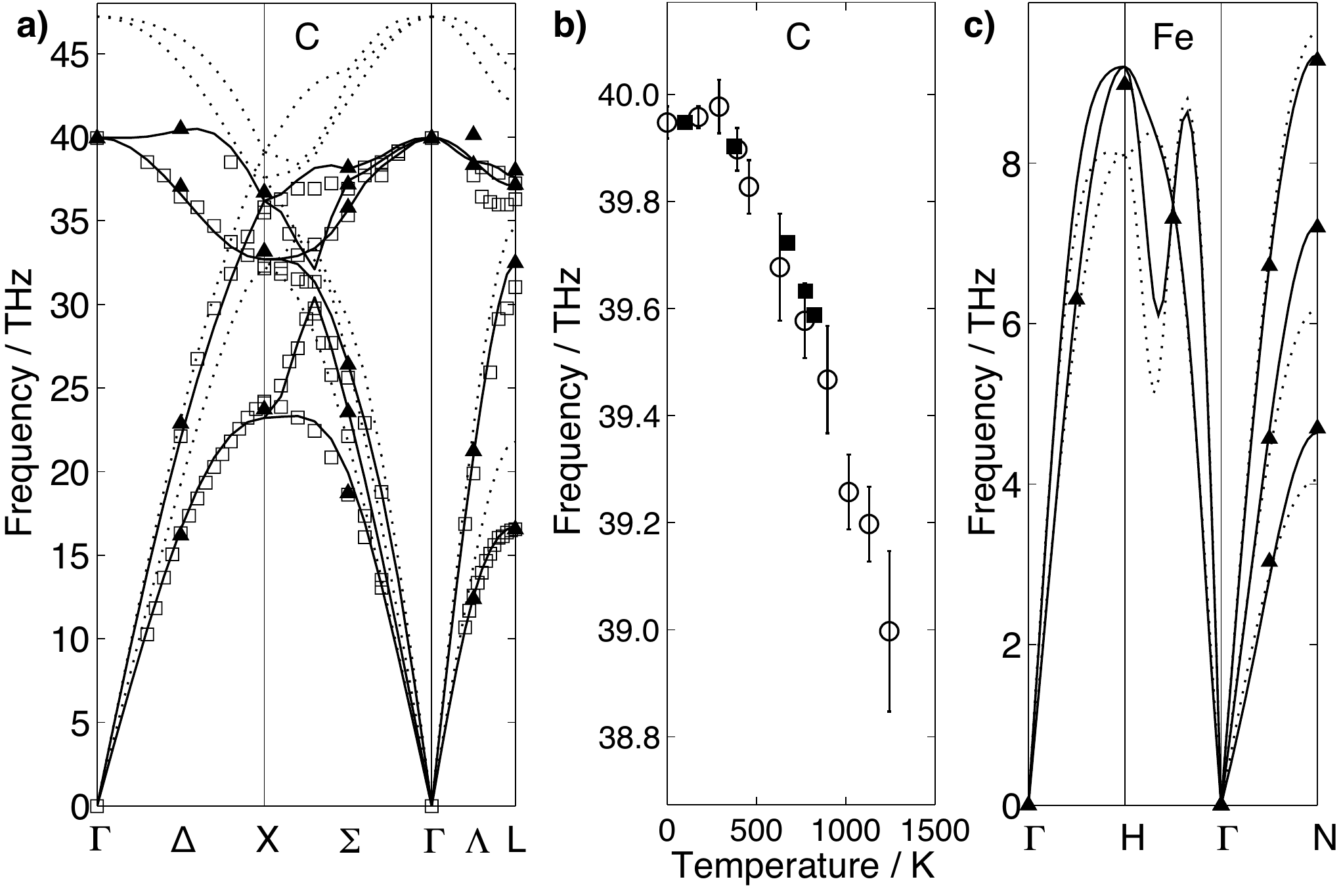}
\caption{\label{fig:phonon} a) Phonon dispersion curves for diamond using the GAP model (lines), DFT
(triangles) and experiment (squares)\cite{phonon12}; b) temperature dependence of the
$\Gamma_{25^\prime}$ mode from MD using GAP (circles, 250 atoms, 20 ps) and experiment
(squares)\cite{phonon09}. In accordance with common practice\cite{phonon18} the calculated points
have been shifted by a constant to agree with experiment at zero temperature to account for the
anharmonic effects of zero-point motion and a quantum correction to the kinetic temperature is also applied\cite{phonon02}; c) phonon dispersion of iron using GAP (solid), Finnis-Sinclair potential \cite{fs01} (dotted) and DFT (triangles). 
} 
\end{figure}

\begin{figure}[th]
\includegraphics[width=8.5cm]{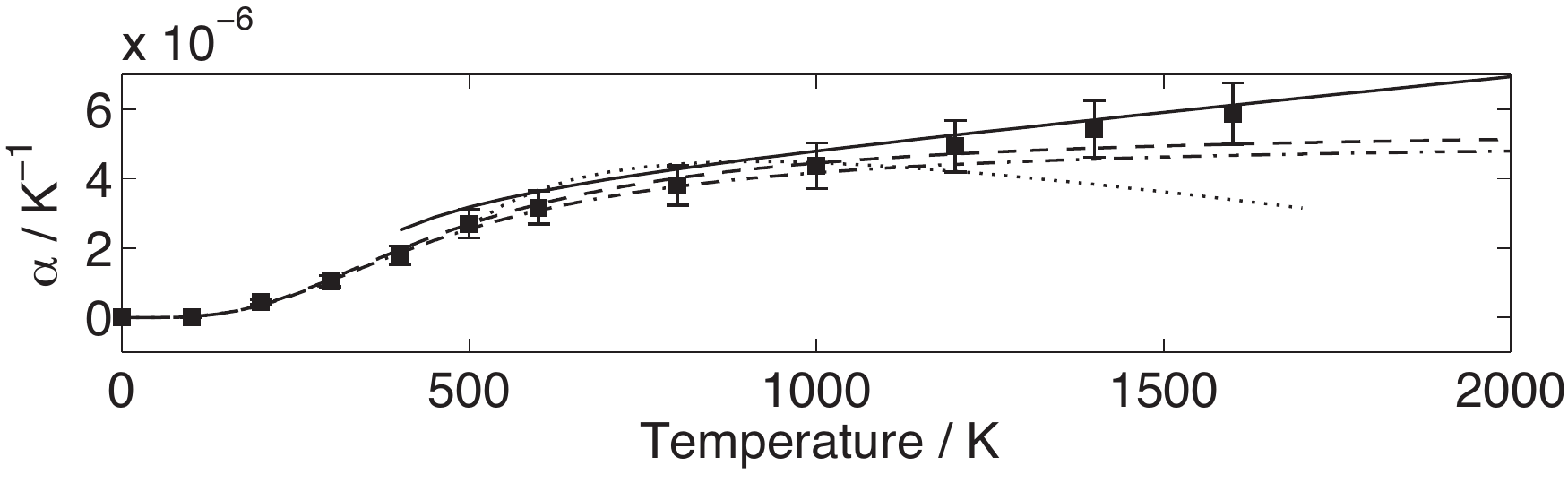}
\caption{\label{fig:thexp} Linear thermal expansion coefficient of
  diamond in the GAP model (dashed) and DFT (dash-dotted) using
  the quasi-harmonic approximation\cite{phonon19}, and derived from MD  (216 atoms, 40 ps) with
  GAP (solid) and the Brenner potential (dotted). Experimental results are shown with
  squares\cite{thexp02}. 
}
\end{figure}

The lower panels of Fig.~\ref{fig:forcecorr} show the performance of the GAP model for semiconductors in terms of the accuracy of forces for near-bulk configurations. For diamond, the GAP model is shown to improve significantly as the cutoff is increased (there is also systematic improvement as $J_\textrm{max}$ is increased, this is shown in the SI). For all three materials the RMS errors in the energy are less than 1 meV/at. 
Table~\ref{tab:elastic} shows the elastic constants. It is remarkable that the existing potentials are not able to reproduce all elastic constants to better than 25\% for any setting of their parameters. Fig.~\ref{fig:phonon} shows the phonon
spectrum for diamond and iron. For diamond, the GAP model shows excellent accuracy at zero temperature over most of the Brillouin zone,
with a slight deviation for optical modes in the $\Lambda$ direction. The agreement with experiment
is also good for the frequency of the Raman mode as a function of temperature. For iron, the agreement with DFT is even better. We also computed the linear thermal expansion coefficient of diamond, shown in Fig.~\ref{fig:thexp}, using two different methods, applicable at low and high temperatures. Our low temperature curve is derived
from the phonon spectrum via the quasi-harmonic approximation and agrees well with the DFT and
experimental results. At higher temperatures higher order anharmonic terms come into play, so we use
molecular dynamics (MD) 
and obtain good agreement with experiment, showing that the GAP model is accurate significantly
beyond the small displacements that control phonons. 



\begin{figure}[th]
\includegraphics[width=8cm]{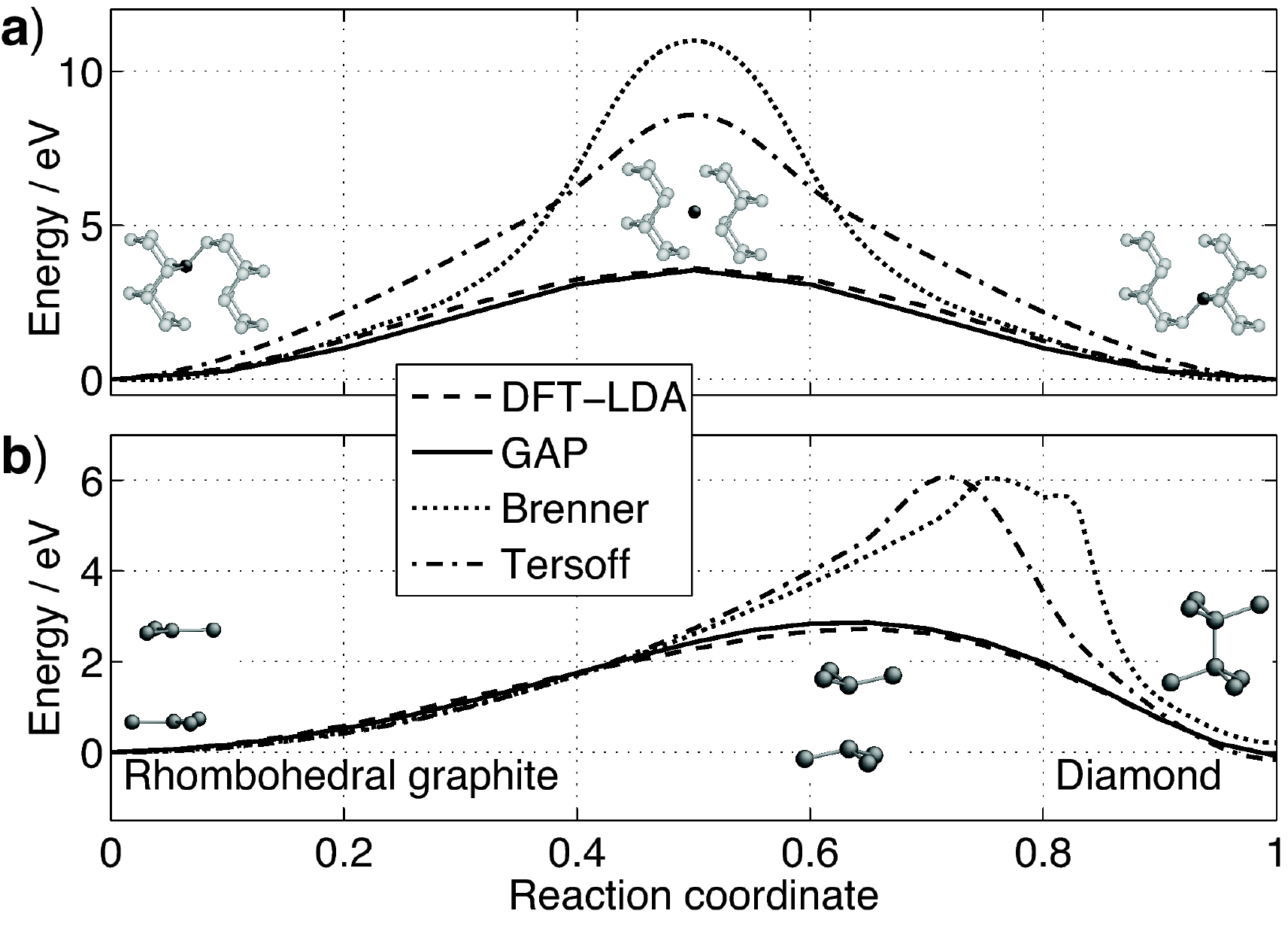}
\caption{\label{fig:gra2dia} The energetics of the linear transition path for a migrating vacancy (top) and for the rhombohedral graphite to diamond transformation.}
\end{figure}

Finally, we extended the GAP model by including reference configurations generated by random
displacements around a diamond vacancy and graphite. Fig.~\ref{fig:gra2dia} shows the energetics of the transition path
for a migrating vacancy in diamond and the transition from rhombohedral graphite to diamond. The agreement with DFT is excellent, demonstrating that we can construct a truly reactive model that describes the $sp^2$-$sp^3$ transition correctly, in contrast to currently used interatomic potentials. 

Even for the small systems considered above, the GAP model is orders of magnitude faster than
standard plane-wave DFT codes, but significantly more expensive than simple analytical potentials.
The computational cost is roughly comparable to the cost of numerical bond order potential
models\cite{Mrovec04}. The current implementation of the GAP model takes 0.01~s/atom/timestep on a
single CPU core.
For comparison, a timestep of the 216-atom unit cell of Fig.~\ref{fig:thexp} takes 191~s/atom using
Castep, about 20,000 times longer, while the same for iron would take more than a {\em million} times longer.

In summary, we have outlined a framework for automatically generating finite range interatomic
potential models from quantum-mechanically calculated atomic forces and energies. The models were
tested on bulk semiconductors and iron and were found to have remarkable accuracy in matching the {\em ab
initio} potential energy surface at a fraction of the cost, thus demonstrating the fundamental
capabilities of the method. Preliminary data for GaN, presented in the SI, shows that the extension to multi-component
and charged systems is straightforward by augmenting the local energy with a simple Coulomb term using fixed charges.    
Our long term goal is to expand the range of interpolated configurations and thus create``general'' interatomic potentials for one- and two-component materials whose accuracy approaches that of quantum mechanics. 

\begin{acknowledgments}
  The authors thank Sebastian Ahnert, Noam Bernstein, Zoubin Ghahramani,
  Edward Snelson and Carl Rasmussen
  for discussions. APB is supported by the EPSRC. GC acknowledges
  support from the EPSRC under grant number EP/C52392X/1. Part of the
  computational work was carried out on the Darwin Supercomputer of
  the University of Cambridge High Performance Computing Service.
\end{acknowledgments}

\bibliography{gap}

\end{document}


\maketitle

\section{Bispectrum}

An arbitrary function $\rho$ defined on the surface of a 4D sphere can be numerically represented
using the hyperspherical harmonic functions $U^j_{m'm}(\phi, \theta, \theta_0)$. The hyperspherical
harmonics form an orthonormal basis set thus $\rho$ can be represented as
\begin{equation*}
\rho = \sum _{j=0} ^\infty \sum _{m,m'=-j}^{j} c^j_{m'm} U^j_{m'm} \textrm{.}
\end{equation*}
The expansion coefficients $c^j_{m'm}$ can be calculated via
\begin{equation*}
c^j_{m'm} = \langle U^j_{m'm} | \rho \rangle \textrm{,}
\end{equation*}
where $\langle . | . \rangle$ denotes the inner product. For clarity, the vectors $\mathbf{c}^j$ are constructed from
the expansion coefficients $c^j_{m'm}$.
A unitary operation $\hat{R}$, such as a rotation, acting on $\rho$ transforms the coefficient
vectors $\mathbf{c}^j$ according to
\begin{equation*}
\mathbf{c'}^{j} = \mathbf{R}^j \mathbf{c}^j \textrm{,}
\end{equation*}
where $\mathbf{R}^j$ are unitary matrices, i.e. $\left( \mathbf{R}^j \right) ^\dagger \mathbf{R}^j = \mathbf{I}$.

The direct product of two rotational matrices, $\mathbf{R}^{j_1}$ and $\mathbf{R}^{j_2}$ can be
decomposed into a direct product of $\mathbf{R}^j$ matrices by a unitary transformation
\begin{equation*}
\mathbf{R}^{j_1} \otimes \mathbf{R}^{j_2} = \left( \mathbf{H}^{j_1,j_2} \right)^\dagger
\left[ \bigoplus_{j=|j_1-j_2|}^{j_1+j_2} \mathbf{R}^{j} \right] \mathbf{H}^{j_1,j_2} \textrm{,}
\end{equation*}
where the matrix $\mathbf{H}^{j_1 j_2}$ is the four-dimensional analouge of the Clebsch-Gordan
coefficients. In fact, the elements of the matrix are obtained as product of Clebsch-Gordan
coefficients: $H^{l m m'}_{l_1 m_1 m_1',l_2 m_2 m_2'} \equiv C^{lm}_{l_1 m_1 l_2 m_2} C^{lm'}_{l_1
m_1' l_2 m_2'}$.
The direct product of the coefficient vectors $\mathbf{c}^{j_1}$ and $\mathbf{c}^{j_2}$ transforms
according to the direct product of the rotational matrices
\begin{equation*}
\mathbf{c}^{j_1} \otimes \mathbf{c}^{j_2} \to \left\{ \mathbf{R}^{j_1} \otimes \mathbf{R}^{j_2} \right\}
\mathbf{c}^{j_1} \otimes \mathbf{c}^{j_2} = \left\{ \left( \mathbf{H}^{j_1,j_2} \right)^\dagger
\left[ \bigoplus_{j=|j_1-j_2|}^{j_1+j_2} \mathbf{R}^{j} \right] \mathbf{H}^{j_1,j_2} \right\}
\mathbf{c}^{j_1} \otimes \mathbf{c}^{j_2}
\textrm{.}
\end{equation*}
We define $\mathbf{g}^{j_1,j_2,j}$---using the fact that $\mathbf{H}^{j_1,j_2}$ is unitary---as follows:
\begin{equation*}
\left[ \bigoplus_{j=|j_1-j_2|}^{j_1+j_2} \right] \mathbf{g}^{j_1,j_2,j} \equiv \mathbf{H}^{j_1,j_2}
\mathbf{c}^{j_1} \otimes \mathbf{c}^{j_2} 
\textrm{,}
\end{equation*}
which transforms under rotation as $\mathbf{g}^{j,j_1,j_2} \to  \mathbf{R}^j
\mathbf{g}^{j_1,j_2,j}$. The cubic rotational invariants, also known as the bispectrum, can be
constructed as
\begin{equation*}
B_{j_1,j_2,j} = \left( \mathbf{c}^j \right)^{\dagger} \mathbf{g}^{j_1,j_2,j}
\textrm{.}
\end{equation*}
Finally, we arrive to the expression for the bispectrum elements, computed as
\begin{equation*}
B_{j_1,j_2,j} = \sum_{m'_1, m_1 = -j_1}^{j_1} \sum_{m'_2, m_2 = -j_2}^{j_2}
\sum_{m', m = -j}^{j}
\bigl( c^j_{m'm} \bigr)^* \\C^{jm}_{j_1 m_1 j_2 m_2} C^{jm'}_{j_1 m'_1 j_2 m'_2}
c^{j_1}_{m'_1 m_1} c^{j_2}_{m'_2 m_2} \textrm{.}
\end{equation*}

The truncated version of the bispectrum results in a finite array, with 4, 23 and 69 elements for
$J_\textrm{max}=1$, 3 and 5, respectively.

\section{Gaussian Process Regression}

\parindent=0pt
\parskip=5pt
Notation and formulae:

\begin{eqnarray*}
N &:& \textrm{Number of raw atomic neighbourhood configurations}\\
M &:& \textrm{Number of sparse atomic neighbourhood configurations}\\
\mathbf{x}_n&:& \textrm{bispectrum of $n$th reference configuration}\\
\overline{\mathbf{x}}_m&:& \textrm{bispectrum of $m$th sparse configuration}\\
\mathbf{x}_* &:& \textrm{bispectrum of configuration for which prediction is sought (``test configuration'')}\\
\mathbf{y} &:& \textrm{vector of data values at the raw configurations}\\
C(\mathbf{x}, \mathbf{x}^\prime) &:& \textrm{Covariance function, the measure of similarity of two configurations}\\
\left[\mathbf{C}_N \right]_{nn'} &=& C(\mathbf{x}_n,\mathbf{x}_{n'}), \textrm{ covariance matrix of raw configurations}\\
\left[\mathbf{C}_M \right]_{mm'} &=& C(\overline{\mathbf{x}}_m,\overline{\mathbf{x}}_{m'}), \textrm{ covariance matrix of sparse configurations}\\
\left[\mathbf{C}_{NM}\right]_{nm} &=& [\mathbf{k}_n]_m = C(\mathbf{x}_n,\overline{\mathbf{x}}_m), \textrm{ covariance matrix of sparse and raw configurations, } \mathbf{C}_{MN} = (\mathbf{C}_{NM})^\textrm T\\
\left[\mathbf{k}_*\right]_m &=& C(\overline{\mathbf{x}}_m,\mathbf{x}_*),  \textrm{ covariance vector of test and sparse configurations}\\
\mathbf{\Lambda} &=& \mathrm{Diag}( \mathrm{diag}(\mathbf{C}_{N} - \mathbf{C}_{NM} \mathbf{C}_M^{-1} \mathbf{C}_{MN}))\\
\sigma  &:& \textrm{intrinsic noise of data values (hyperparameter)}\\
\mathbf{Q}_M &=& \mathbf{C}_M + \mathbf{C}_{MN}
( \mathbf{\Lambda} + \sigma^2 \mathbf{I})^{-1} \mathbf{C}_{NM}, \textrm{ pseudo-covariance matrix of the sparse configurations}\\
\varepsilon_* &=& \mathbf{k}_*^T \mathbf{Q}_M^{-1} \mathbf{C}_{MN}
( \mathbf{\Lambda} + \sigma^2 \mathbf{I})^{-1} \mathbf{y}, \textrm{ prediction of atomic energy for test configuration}\\
\sigma^2_* &=& C(\mathbf{x}_*,\mathbf{x}_*) - \mathbf{k}_*^T
(\mathbf{C}_M^{-1}-\mathbf{Q}_M^{-1}) \mathbf{k}_* + \sigma^2, \textrm{ variance of prediction for test configuration}
\end{eqnarray*}

where $\mathrm{diag}(\mathbf{A})$ is the vector of diagonal elements of the matrix $\mathbf{A}$, and $\mathrm{Diag}(\mathbf{v})$ is the matrix whose diagonal elements are the components of vector $\mathbf{v}$ and the off-diagonal elements are zero. 

%
%

\subsection{Covariance function (kernel)}

We use a Gaussian kernel. This kernel enables us to assign a separate {\em spatial scale} hyperparameter ($\theta_i$) to each element of the feature vector, therefore the kernel provides a more flexible description than a Gaussian kernel with a single hyperparameter. The next three equations show how the covariance function is evaluated if the function values are available for both configurations, if we have the derivative of the function available at one configuration, and if the derivatives are given for both configurations.

\begin{eqnarray*}
C(\mathbf{x}_n,\mathbf{x}_m) &=& \delta^2 \exp \left( -\frac{1}{2} \sum_i \frac{(x_n^i -
x_m^i)^2}{\theta_i^2} \right)\\
C^\prime(\mathbf{x}_n,\mathbf{x}_m) &=& \delta^2 \exp \left( -\frac{1}{2} \sum_i \frac{(x_n^i -
x_m^i)^2}{\theta_i^2} \right) \sum_i \frac{x_m^i - x_n^i}{\theta_i^2} \frac{\partial x^i_m}{\partial
r_\alpha}\\
C^{\prime\prime}(\mathbf{x}_n,\mathbf{x}_m) &=& \delta^2 \exp \left( -\frac{1}{2} \sum_i \frac{(x_n^i -
x_m^i)^2}{\theta_i^2} \right) \left[ \sum_i \frac{1}{\theta_i^2} \frac{\partial x^i_n}{\partial
r_\alpha} \frac{\partial x^i_n}{\partial r_\beta} - \left( \sum_i \frac{x_n^i - x_m^i}{\theta_i^2}
\frac{\partial x^i_n}{\partial r_\alpha} \right)
\left( \sum_i \frac{x_n^i - x_m^i}{\theta_i^2}\frac{\partial x^i_m}{\partial r_\beta} \right) \right]\\
\end{eqnarray*}

\subsection{Linear combinations}

In our case, only the linear combination of atomic energies can be directly observed. We cannot determine the atomic contributions to the total energy of a system uniquely from an electronic structure calculation. Similarly, the atomic forces---although they are available from first-principles calculations---are not derivatives of atomic energies, but are sums of derivatives of different atomic contributions. It is possible to use a Gaussian Process to infer the underlying function even if only linear combinations of function values are available. Now let $\mathbf{y}$ is the vector of $K$ observed values (total energies and atomic force components). Let $\mathbf{y}^\prime$ be the vector of $N$ {\em unobserved} values of atomic energies and its derivatives corresponding to the $N$ atomic neighborhood configurations. Let the $N\times K$ matrix $\mathbf{L}$ describe the relationship of the $K$ observations to the $N$ unknown values. The elements of $\mathbf{L}$ are 0s and 1s, and 

\[
\mathbf{y} = \mathbf{L}\mathbf{y}'
\]
The covariance of the $K$ observations is then given by
\[ \mathbf{C}_{KK} = \mathbf{L}^\mathrm T \mathbf{C}_{NN} \mathbf{L} \].

\subsection{Putting it all together}

The sparsification and the linear combinations are used together to give the final expression for the atomic energy, in such a way that the unobserved values $\mathbf{y}^\prime$ are not needed,

\begin{eqnarray*}
\mathbf{\Lambda} &=& \mathrm{Diag}\left(  \mathrm{diag}( \mathbf{L}^T
\mathbf{C}_{NN} \mathbf{L} -
\mathbf{L}^T \mathbf{C}_{NM} \mathbf{C}_{M}^{-1} \mathbf{C}_{MN}
\mathbf{L} ) \right)\quad (\textrm{now a } K\times K \textrm{ diagonal matrix})\\
\varepsilon_* &=& \mathbf{k}_*^T \left[ \mathbf{C}_M + \mathbf{C}_{MN}
\mathbf{L} ( \mathbf{\Lambda} +
\sigma^2 \mathbf{I})^{-1} \mathbf{L}^T \mathbf{C}_{NM} \right]^{-1}
\mathbf{C}_{MN} \mathbf{L}
( \mathbf{\Lambda} + \sigma^2 \mathbf{I})^{-1} \mathbf{y} \\
\end{eqnarray*}

\section{The data: density functional theory}

The DFT data was generated using the local density approximation in case of carbon and the PBE generalized gradient approximation for silicon, germanium, GaN and iron. The electronic Brillouin zone was sampled by using a Monkhorst-Pack k-point grid, with a k-point spacing of at most $0.3 \mathrm{\AA}^{-1}$ for insulators and $0.14 \mathrm{\AA}^{-1}$ for iron. The plane wave cutoff was 350, 300, 300, 350, 500 eV for C, Si, Ge, GaN and Fe respectively and the energies were extrapolated to correct for the finite basis set. Ultrasoft pseudopotentials were used with 4 valence electrons for all group IV ions, 3 electrons for Ga, 5 electrons for N and 8 for Fe ions.

\section{Testing the GAP parameters}

\begin{figure}
\begin{center}
\includegraphics[width=10cm]{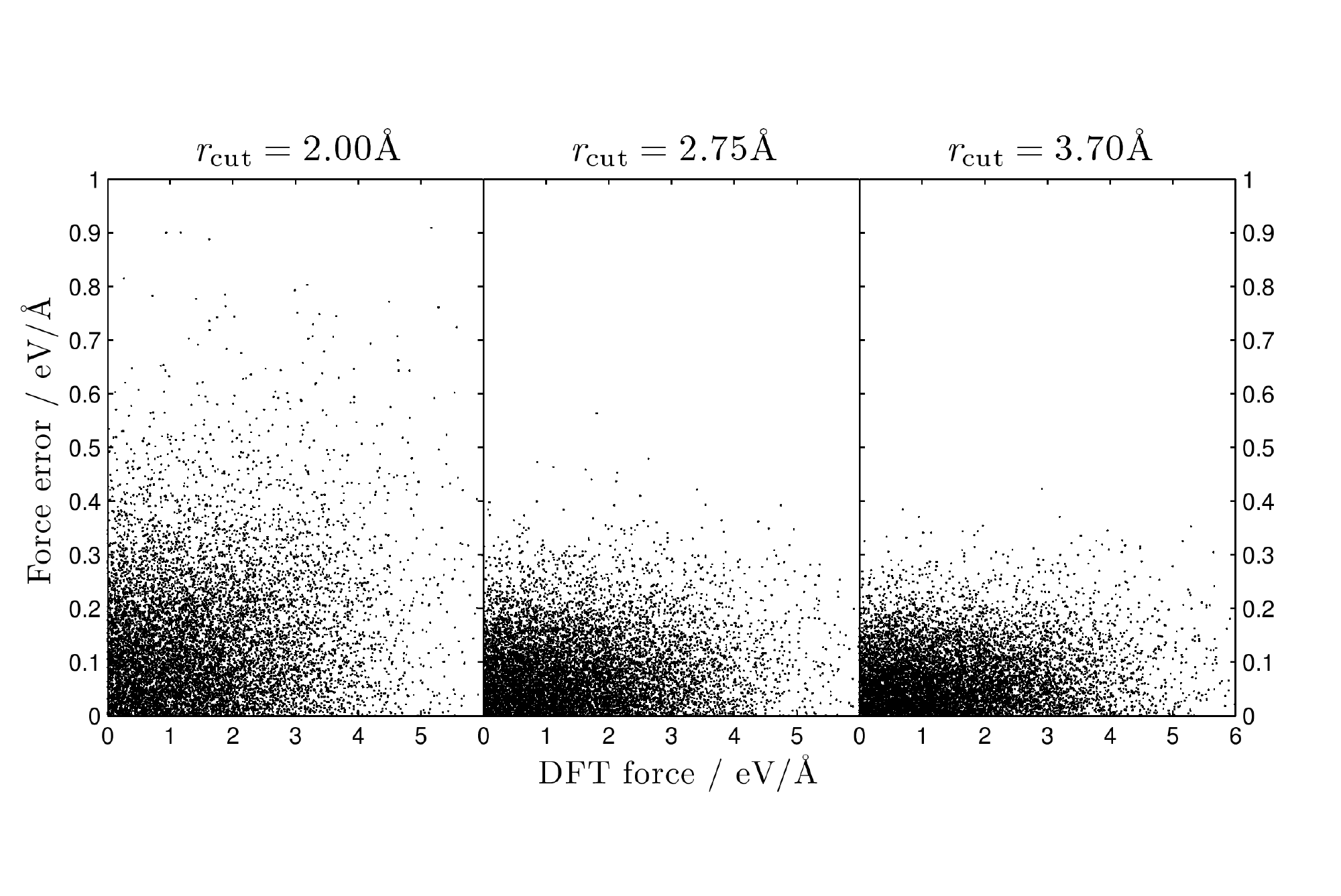}
\end{center}
\caption{\label{fig:rcut} Force correlation of GAP models for diamond with different
spatial cutoffs.}
\end{figure}

\begin{figure}
\begin{center}
\includegraphics[width=10cm]{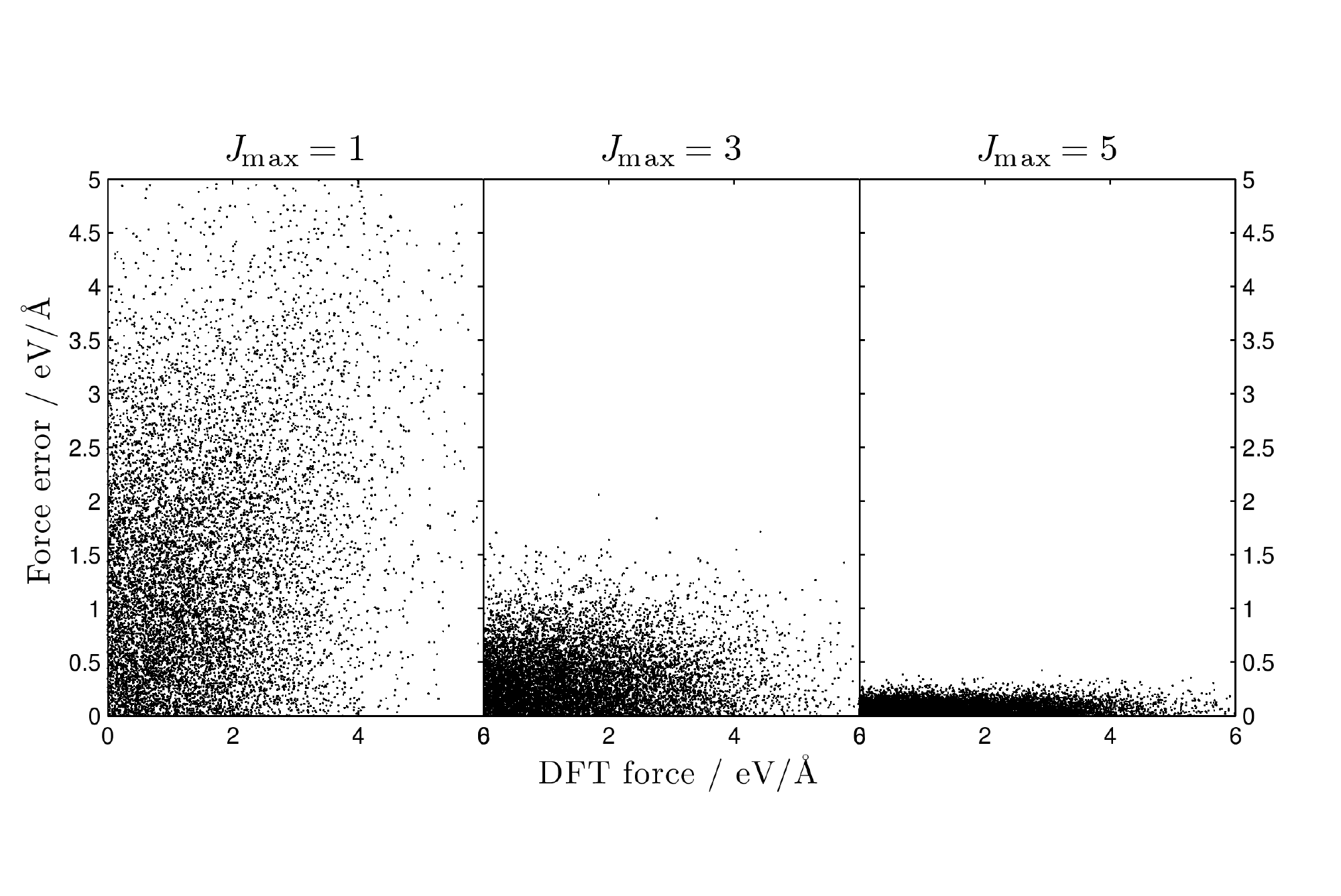}
\end{center}
\caption{\label{fig:J} Force correlation of GAP models for diamond with different resolution of
representation. The number of invariants were 4, 23 and 69 for $J_{\textrm{max}}=$1, 3 and 5,
respectively.}
\end{figure}

Figure~\ref{fig:rcut} shows the improvement in the distribution of force errors of the GAP model for diamond as the cutoff radius is increased. Figure~\ref{fig:J} shows the same as $J_\mathrm{max}$ is increased, which corresponds to increasing the spatial resolution of the bispectrum.

\section{Potential for gallium nitride}

\begin{figure}
\begin{center}
\includegraphics[width=10cm]{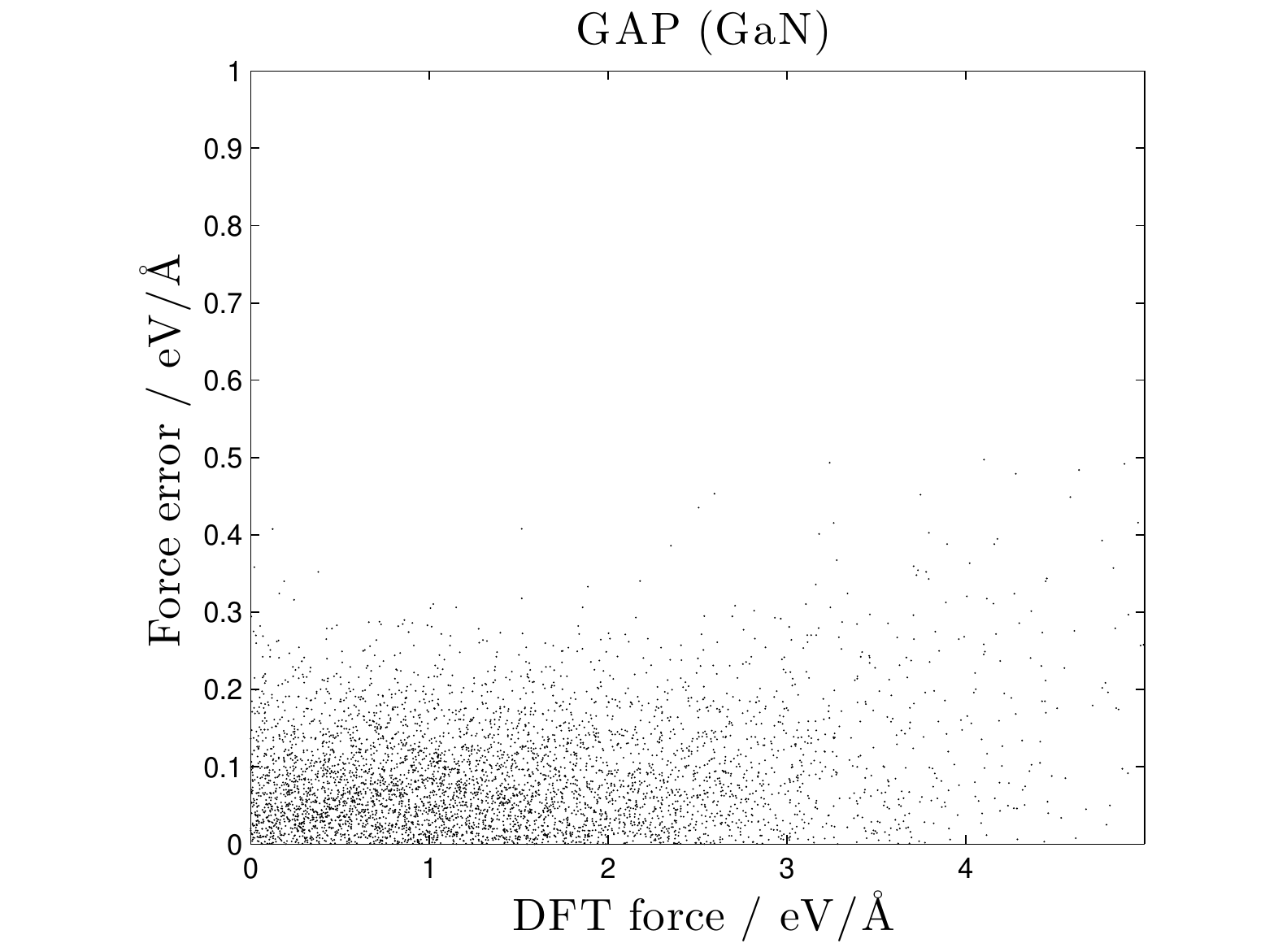}
\end{center}
\caption{\label{fig:corr_gan} Force errors in GaN of the GAP model augmented with a simple Ewald sum of fixed charges with reference to DFT forces.}
\end{figure}

\begin{figure}
\begin{center}
\includegraphics[width=10cm]{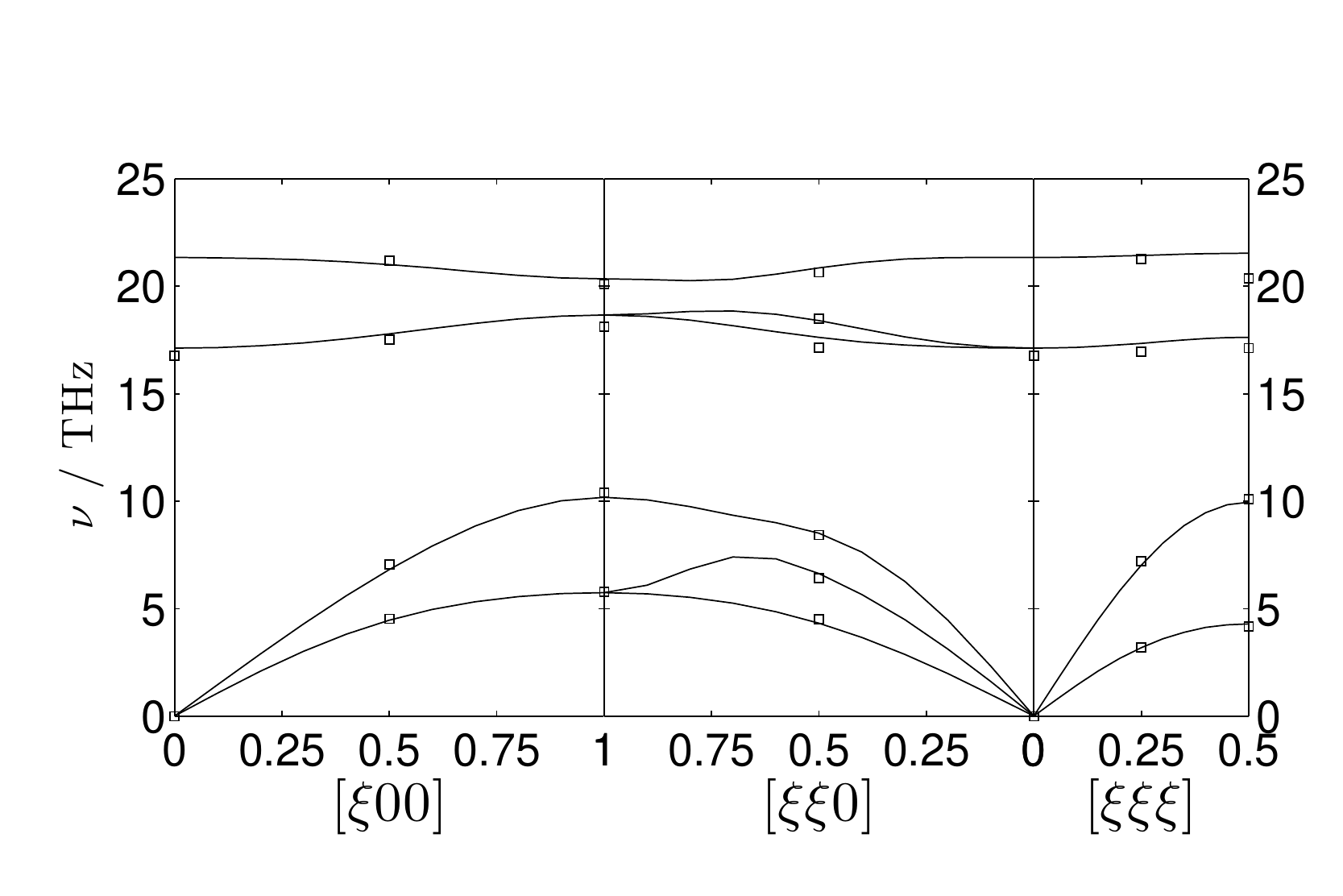}
\end{center}
\caption{\label{fig:gan_phonon} Phonon spectrum of GaN calculated with GAP (solid lines) and PBE-DFT
(open squares).}
\end{figure}

Figures~\ref{fig:corr_gan} and \ref{fig:gan_phonon} show the force errors and the phonon spectrum of a simple GAP model for gallium nitride. The long range Coulomb interactions of the ions is significant in this system, so we augmented the local energy of the original GAP model by an Ewald sum of fixed charges (+1 for Ga and -1 for N). The Gaussian Process regression was carried out on forces and energies which were obtained from the DFT calculations by subtracting this Coulomb contribution. The LO/TO splitting in the phonon spectrum shows that the model captures the long range character of the ionic interactions correctly.

